\documentclass[aps,prb,showpacs,preprintnumbers,twocolumn,superscriptaddress]{revtex4-2}
\usepackage{amsmath,amssymb}
\usepackage{bm}
\usepackage{tipa}
\usepackage{upgreek}
\usepackage{comment}
\usepackage{mathrsfs}
\usepackage{graphicx}
\usepackage{braket}
\usepackage{enumitem}
\usepackage{mathbbol}
\usepackage{booktabs}
\usepackage{gensymb}
\usepackage[normalem]{ulem}
\usepackage{color}
\usepackage[colorlinks,bookmarks=true,citecolor=blue,linkcolor=red,urlcolor=blue]{hyperref}
\usepackage{hyperref}

\usepackage{pifont}

\allowdisplaybreaks

\usepackage{siunitx}
\usepackage{soul}

\begin{document}

	\title{Yu-Shiba-Rusinov tips:\\ imaging spins at the atomic scale with full magnetic sensitivity
} 
	
	\author{Felix K{\"u}ster$^1$, Souvik Das$^1$, Stuart S.\ P.\ Parkin$^1$, Paolo Sessi}
	
	\affiliation{
	Max Planck Institute of Microstructure Physics, Halle (Saale) 06120, Germany,\\
	}

	\begin{abstract}
    Measurements of magnetic properties at the atomic scale require probes capable of combining high spatial resolution with spin sensitivity. Spin-polarized scanning tunneling microscopy (SP-STM) fulfills these conditions by using atomically sharp magnetic tips. The imaging of spin structures results from the tunneling magneto-conductance that depends on the imbalance in the local density of spin-up and spin-down electrons. Spin-sensitive tips are generally formed from bulk materials or by coating non-magnetic tips with a thin magnetic layer. However, ferromagnetic materials generate stray magnetic fields which can influence the magnetic structure of the probed system, while the magnetization of antiferromagnetic materials is difficult to set tip by externally applied magnetic fields. Here, we use functionalized Yu-Shiba-Rusinov (YSR) tips prepared by attaching magnetic adatoms at the apex of a superconducting cluster to image magnetic interactions at the atomic scale. We demonstrate that YSR tips are capable of sensing different magnetization directions, conferring them full magnetic sensitivity. We additionally show that the finite size of the tip superconducting cluster makes it robust against relatively strong  magnetic fields, making YSR tips capable of visualizing magnetic field driven transitions of the spin texture. 
	\end{abstract} 
 	\maketitle
 
\textbf{Introduction.}
Spin-polarized electron tunneling exploits the tunneling magnetoresistance between two spin-polarized electrodes to access the respective spin-dependent electronic densities of states \cite{MESERVEY1994173,PhysRevB.39.6995}. This phenomenon has been used to study the spin-dependent properties of various types of materials including magnetic metals \cite{JULLIERE1975225}, semiconductors \cite{PRINS1993152}, and superconductors \cite{PhysRevLett.25.1270}. Using a spin-polarized tip as one of the electrodes in a scanning tunneling microscope (STM) led to the development of spin-polarized STM (SP-STM) \cite{PhysRevLett.65.247}, a technique which allows one to correlate the structural, electronic, and magnetic properties of surfaces with atomic resolution \cite{MBode_2003,RevModPhys.81.1495}.  SP-STM has significantly evolved over time, providing valuable insights into several magnetic phenomena such as the formation of magnetic domains \cite{PhysRevLett.87.127201}, the emergence of complex non-collinear magnetic textures \cite{BHB2007,HBM2011}, the visualization of RKKY-mediated interactions \cite{ZWL2010}, and the detection of elementary spin excitations in single atoms \cite{doi:10.1126/science.1101077,PhysRevLett.106.037205} as well as their evolution in artificially crafted nanostructures \cite{doi:10.1126/science.1125398}. SP-STM is nowadays a mature technique, where the spin-dependent signal contributing to the differential conductance $dI/dU$ at the energy $eU$ with respect to the Fermi level $E_F$ is generally modeled as: 
$$dI/dU \propto \overrightarrow{m}_T\overrightarrow{m}_S(E_F + eU) $$
where $\overrightarrow{m}_T$,$\overrightarrow{m}_S$ are vectors corresponding to the spin polarized local density of stats (LDOS) in tip and sample, respectively. 

SP-STM tip are generally prepared using magnetic bulk tips or by coating electrochemically-etched W tip with thin films of magnetic elements. Besides spin-sensitivity, the ideal SP-STM tip needs to fulfill additional conditions, namely: (i) negligible stray fields to avoid dipolar interactions influencing or even destroying the sample’s intrinsic magnetic structure \cite{PhysRevLett.88.057201}, (ii) possibility to control the tip magnetization direction in-situ to make it sensitive to distinct magnetic orientations, i.e. in-plane and out-of-plane \cite{PhysRevLett.88.057201}. 
\begin{figure*}[!t]
    \centering
    \includegraphics[width=.75\textwidth,page=1]{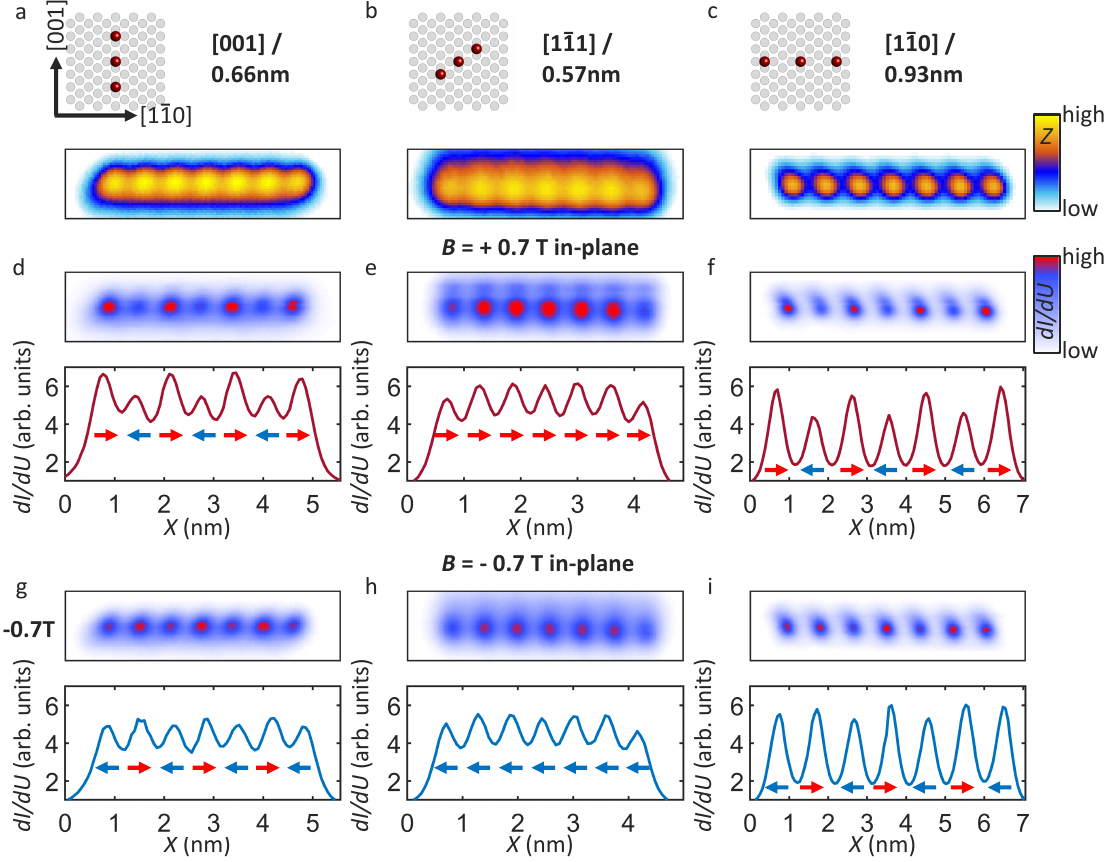}
    \caption{\textbf{Spin-resolved measurements with bulk magnetic tips}
    \textbf{a-c}. Topographic images of three spin chains, each once consisting of seven adatoms, assembled along different crystallographic directions. Top panels report a schematic representation of the chains, showing their orientation as well as the distance between adatoms. Red dots correspond to Cr adatom, grey dots identify the Nb adatom of the underlying Nb(110) substrates.{d-f} d$I$/d$U$ maps and relative line profiles acquired in constant height mode in a magnetic field of $B$ = 0.7 T applied parallel to the sample surface. {g-i} d$I$/d$U$ maps and relative line profiles by reverting the direction of the magnetic field $B$ = -0.7 T. Stabilization parameters $-5~\mathrm{mV}$, $3~\mathrm{nA}$. Spin resolved d$I$/d$U$ maps: scanning bias $0.27~\mathrm{mV}$, modulation $80~\mathrm{\upmu V}$.}
    \label{fig:Figure1}
\end{figure*}
  
  Conditions (i) and (ii) are generally difficult to be simultaneously fulfilled. On the one hand, the use of ferromagnetic materials as tip coating layers allows to controllably orient the tip magnetization in applied magnetic fields, but it inevitably results in tip-induced stray fields.  On the other hand, the use of antiferromagnetic materials  allows to avoid stray fields but it impedes to control the tip magnetization direction due to the absence of a net magnetic moment.  
  
  Since the tunnelling current decays exponentially with increasing distance, a single magnetic atom or molecules localized at the tip very end is expected to generate a sizable spin-dependent tunneling signal. This consideration has been used to demonstrate how spin-sensitive measurements can be obtained using a detection scheme which relies on the spin-excitations of single magnetic molecules attached to a conventional STM tip \cite{doi:10.1126/science.aax8222,doi:10.1126/science.aaw7505}.  More recently, it has been reported that high-spin sensitivity can by achieved by using superconducting tip functionalized with magnetic atoms at its apex. The general idea is to use the spin-polarized Yu-Shiba-Rusinov (YSR) states induced by the magnetic exchange of 3$d$ atoms with a superconducting tip to image spin-structures at the atomic scale \cite{doi:10.1126/sciadv.abd7302}. Here, we report on how this methodological approach can be used to create YSR tips capable of (i) sensing in-plane, out-of-plane, and tilted magnetization directions and (ii) imaging magnetic field driven transitions in the spin texture. 
  
  \textbf{Experimental set-up.} Experiments were performed under ultra-high vacuum conditions ($p < 3\cdot10^{-10}$ mbar) in a STM setup operated at a base temperature of $T$ = 550 mK. As a prototypical system, we focus on Cr adatoms coupled to a clean Nb(110) surface \cite{KMG2021,PhysRevB.99.115437}. This choice is motivated by the possibility of creating superconducting tips by indentation of electrochemically etched W tips into the Nb single crystal \cite{doi:10.1126/sciadv.abd7302}. Moreover, the high anisotropy characterizing the Nb(110) Fermi surface \cite{PhysRevB.99.115437,PhysRevB.102.174502} allows, for adatoms indirectly coupled by the Ruderman-Kittel-Kasuya-Yosida (RKKY) interaction, to create Cr nanostructures crafted atom-by-atom which posses distinct magnetic configurations\cite{KBL2021,doi:10.1073/pnas.2210589119}. Finally, the low magnetic anisotropy characterizing Cr adatoms coupled to Nb(110) {\cite{KMG2021} makes possible to controllably act on the nanostructure's magnetization direction, i.e. in-plane vs. out-of-plane by means of externally applied magnetic fields. Further details on sample preparation and STM measurements are reported in Supplementary Information.
  
\textbf{Spin-resolved measurements with bulk magnetic tips.} Figure~\ref{fig:Figure1}\textcolor{red}{a-c} report topographic images of three different Cr chains assembled along three distinct crystallographic directions and characterized by different spacing $a$ between the adatoms, namely: $[001]$ ($a$ = 0.66 nm) , $[1\overline{1}1]$ ($a$ = 0.57 nm), and $[1\overline{1}0]$ ($a$ = 0.93 nm). All chains consist of seven Cr adatoms. An odd number of adatoms guarantees, also in case of antiferromagnetic coupling, the existence of an uncompensated magnetic moment which interacts with an externally applied magnetic field. Figure~\ref{fig:Figure1}\textcolor{red}{d-j} report spin-resolved STM images acquired with an antiferromagnetic Cr bulk tip \cite{doi:10.1063/1.2800810,doi:10.1063/1.3474659}. To stabilize the magnetic configuration against fluctuations, measurements are performed under infulence of an in-plane magnetic field of strength $B$ = 0.7 T. To exclude set-point effects, all spin-resolved d$I$/d$U$ maps have been acquired in the constant height mode with the tunneling junction stabilized on the bare Nb(110) substrate. On the one hand, the d$I$/d$U$ maps and the corresponding line profiles acquired on chains oriented along the $[001]$ (Fig.~\ref{fig:Figure1}\textcolor{red}{d}) and $[1\overline{1}0]$ (Fig.~\ref{fig:Figure1}\textcolor{red}{f}) directions are both characterized by an alternating contrast, an observation indicating an antiferromagnetic coupling between adjacent adatoms. On the other hand, the wire oriented along the $[1\overline{1}1]$ direction (Fig.~\ref{fig:Figure1}\textcolor{red}{e}) shows a constant signal inside the chain, an observation consistent with the existence of ferromagnetic coupling between adatoms (a different signal is observed for the two adatoms localized at the opposite ends of the chain, possibly due to their different local environment, resulting in a different electronic contribution to the local density of states). 

To confirm this hypothesis, the same set of measurements have been repeated by reverting the direction of magnetic field ($B$ = - 0.7 T). This is expected to revert the contrast across the chains due to the existence on an uncompensated moment which allows them to paramagnetically align with the field, while the antiferromagnetic bulk structure of the tip is left unaffected. As illustrated in Fig.~\ref{fig:Figure1}\textcolor{red}{g-j}, the magnetic contrast is reverted for all chains. These observations confirm the existence of antiferromagnetic coupling between adjacent adatoms for the chains oriented along $[001]$ and $[1\overline{1}0]$ direction and ferromagnetic coupling for the chain oriented along the $[1\overline{1}1]$. We would like to note that, in all cases, the experimental results are in agreement with the magnetic coupling along distinct directions as theoretically predicted by $ab-initio$ calculations \cite{KBL2021}.

\textbf{Preparation of YSR tips.}

Starting from the characterization of the spin coupling along different chains obtained using  conventional SP-STM tips, we probe similar structures using functionalized superconducting tips. These tips are produced following a two steps procedure, namely: (i) indentation into the Nb(110) substrate to create a tip whose apex consists of a Nb cluster; (ii) attachment of single Cr adatoms at the tip apex by atomic manipulation pick-up techniques. A similar approach using Fe and Mn adatoms has been reported in Ref.\onlinecite{doi:10.1126/sciadv.abd7302}. Figure~\ref{fig:Figure2}\textcolor{red}{a,b} show representative spectra acquired after each preparation step. Figure~\ref{fig:Figure2}\textcolor{red}{a} reports d$I$/d$U$ curves acquired after step (i). The gray curve corresponds to the spectrum measured without magnetic field. Its lineshape corresponds to the typical convoluted spectrum of superconducting tip and sample density of states ~\cite{doi:10.1126/science.1202204}. It is characterized by single particle coherence peaks energetically located at approximately 3 meV, i.e. twice the size of the Nb superconducting energy gap. The blue curve reports a typical spectrum acquired when an external magnetic field is applied to quench the sample superconducting state. A clear gap is still visible at the Fermi level. This corresponds to the residual tip superconducting gap. Its robustness against magnetic fields is a direct consequence of the finite size superconducting cluster attached to the tip apex \cite{PhysRevLett.25.1270,PhysRevB.77.054511,nl5037947}. Figure~\ref{fig:Figure2}\textcolor{red}{b} reports d$I$/d$U$ curves acquired after step (ii), i.e. when the superconducting tip has been functionalized by pick-up of single Cr adatoms. The gray curve corresponds to the spectrum acquired without magnetic field.  Clear peaks are visible inside the superconducting gap. These correspond to the so-called YSR states, which are induced by the magnetic exchange between magnetic impurities coupled to a superconducting host \cite{Yu,Shiba,Rusinov}. The blue curve corresponds to the spectrum acquired once a magnetic field is applied to quench the substrate superconducting state. A dip corresponding to the tip superconducting gap is still visible, which is superimposed to a step-like function at negative energy (see Supplementary Information for a description of the spectral lineshape). A clear shoulder is visible inside the gap at sample bias = 0.33 mV, corresponding to a YSR state. Being YSR states spin-polarized \cite{PhysRevLett.119.197002}, tunneling at this energy is expected to grant direct access to the spin coupling between adatoms.

\begin{figure}[!t]
    \centering
    \includegraphics[width=.47\textwidth,page=1]{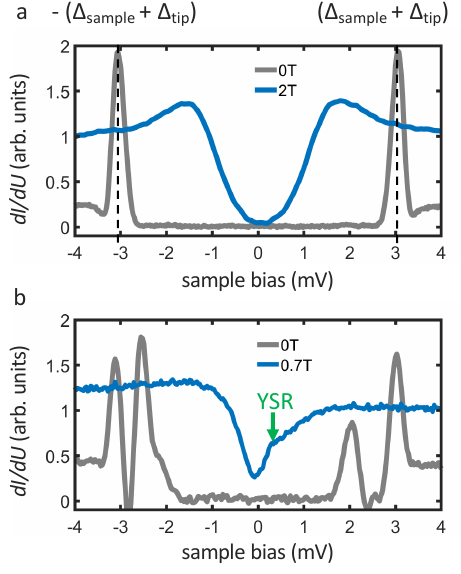}
    \caption{\textbf{Preparation of Yu-Shiba-Rusinov tips.}
    \textbf{a} d$I$/d$U$ curves acquired with a Nb superconducting tip on Nb(110). Gray and blue curves correspond to spectra acquired at $B$ = 0 T and $B$ = 2 T, respectively. A clear superconducting gap is still visible at  $B$ = 2 T, corresponding to the tip superconducting gap. It originates from the small size of the superconducting cluster attached at the tip apex, which results in critical fields significantly higher than in bulk Nb. \textbf{b} d$I$/d$U$ curves acquired at $B$ = 0 T (grey line) and $B$ = 0.7 T (blue line) after a Cr adatom has been attached to the tip superconducting apex. Clear peaks are visible inside the superconducting gap at $B$ = 0 T, corresponding to YSR states. When superconductivity is quenched in the substrate ($B$ = 0.7 T), a YSR state is still visible inside the tip superconducting gap (see shoulder highlighted by a green arrow).}
    \label{fig:Figure2}
\end{figure} 

\textbf{YSR tips with in-plane sensitivity.}

\begin{figure*}[t!]
    \centering
    \includegraphics[width=\textwidth]{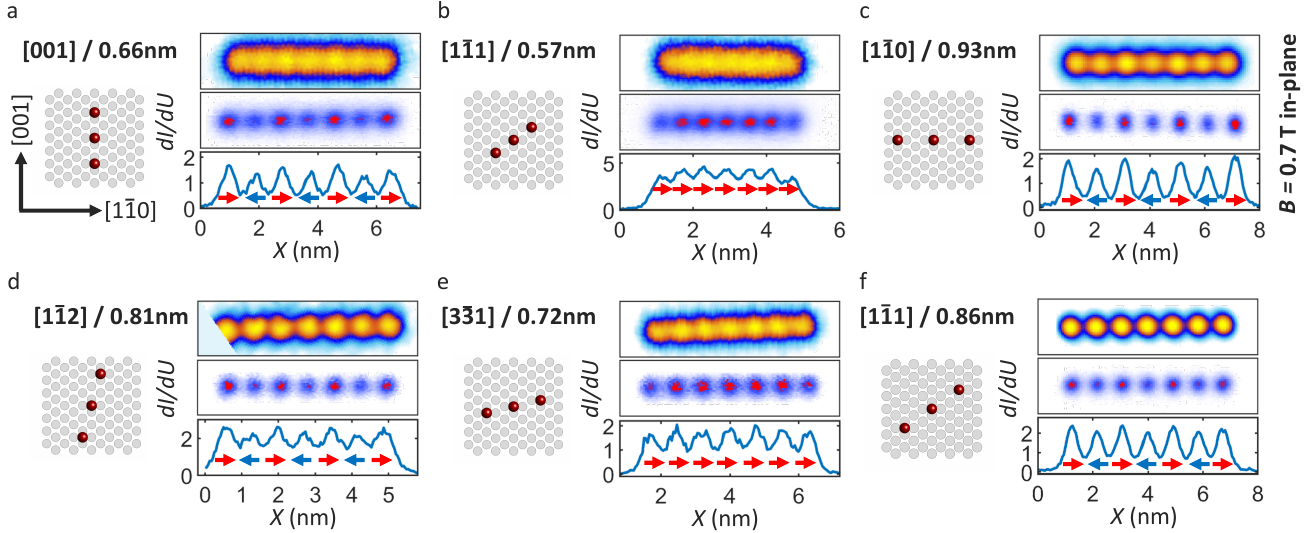}
    \caption{\textbf{Yu-Shiba-Rusinov tips with in-plane sensitivity}
    \textbf{a-f}. Topographic images (top panels), d$I$/d$U$ maps (middle panels) and respective line profiles (bottom panels) for different spin chains assembled along distinct crystallographic directions and by changing the spacing between the adatoms (side panels). Measurements have been performed under a magnetic field $B$ = 0.7 T applied parallel to the sample surface using a YSR tip similar to the one reported in Figure 2, in constant height mode at the bias of the YSR state. Stabilization parameters $-3~\mathrm{mV}$, $500~\mathrm{pA}$, scanning bias $0.33~\mathrm{mV}$, modulation $80~\mathrm{\upmu V}$}
    \label{fig:Figure3}
\end{figure*}

Figure~\ref{fig:Figure3}\textcolor{red}{a-c} reports topographic images,  d$I$/d$U$ maps and corresponding line profiles for chains structurally identical to those scrutinized in Figure~\ref{fig:Figure1}. Measurements have been performed under identical experimental conditions, i.e. under the influence of an in-plane magnetic field of strength $B$ = 0.7 T. All d$I$/d$U$ maps have been acquired in the constant height mode using a YSR tip similar to the one discussed in Figure 2 with the tunneling junction biased at the YSR state, i.e. $V$ = 0.33 mV.  For chains oriented along $[001]$ and $[1\overline{1}0]$ directions, the experimental data reproduce the same alternating antiferromagnetic contrast between
neighboring atoms revealed in Fig.~\ref{fig:Figure1} using a Cr bulk tip. These data demonstrate a strong in-plane magnetic sensitivity of the Yu-Shiba-Rusinov tip. 

It is worth noting that the creation of magnetic nanostructures crafted atom-by-atom requires working with demanding set points, i.e. small biases and high currents, which frequently result in changes in the tip final apex and consequentely loss of its magnetic sensitivity. In this context, the use of Yu-Shiba-Rusinov tip offers a significant advantage: it allows to first create all desired nanostructures and subsequently functionalize the tip in-situ by adatom pick-up techniques, allowing to scrutinize different nanostructures with the very same tip. This is demonstrated in Figure~\ref{fig:Figure3}\textcolor{red}{d-f} which scrutinized the magnetic state of chains assembled by progressively increasing the distance between adatoms and along distinct crystallographic directions. Chains oriented along the $[1\overline{1}2]$ and $[3\overline{3}1]$ are characterized by antiferromagnetic and ferromagnetic coupling between adjacent adatoms, respectively. The comparison of the two chains assembled along the $[1\overline{1}1]$ direction by increasing the spacing, i.e. $a$ = 0.57 nm (Fig.~\ref{fig:Figure3}\textcolor{red}{b}) and $a$ = 0.86 nm (Fig.~\ref{fig:Figure3}\textcolor{red}{f}) evidences a transition from a ferromagnetic to an antiferromagnetic state, an observation in line with the oscillating behavior of RKKY-mediated interactions. 

\textbf{YSR tips with out-of-plane sensitivity.}

\begin{figure}[!t]
    \centering
    \includegraphics[width=.3\textwidth,page=1]{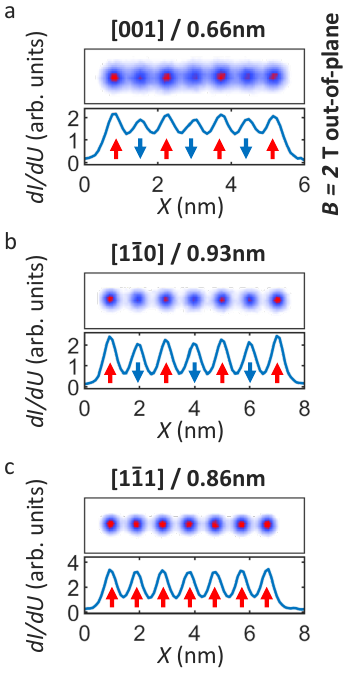}
    \caption{\textbf{Yu-Shiba-Rusinov tips with out-of-plane sensitivity}
    \textbf{a-c}. d$I$/d$U$ (top panels) and corresponding line profiles (bottom panels) acquired on the three antiferromagnetic chains scrutinized in Figure 3, i.e. $[001]$ (spacing $a$ = 0.66 nm) , $[1\overline{1}0]$ ($a$ = 0.96 nm), and $[1\overline{1}1]$ ($a$ = 0.86 nm). The measurements have been acquired under a magnetic field $B$ = 2 T applied perpendicular to the sample surface, using the very same tip and scanning energy as in Figure 3. Chains oriented along $[001]$ and $[1\overline{1}0]$ directions pertain their antiferromagnetic order while a magnetic field driven transition into a ferromagnetic state is visible for the $[1\overline{1}1]$ chain. Stabilization parameters $-3~\mathrm{mV}$, $500~\mathrm{pA}$, scanning bias $0.33~\mathrm{mV}$, modulation $80~\mathrm{\upmu V}$.}
    \label{fig:Figure4}
\end{figure}

YSR tips prepared by single adatoms are characterized by a small magnetic anisotropy \cite{KMG2021}. Therefore, they are expected to easily respond to externally applied magnetic fields, an effect which can be used to controllably make them sensitive to either an in-plane or an out-of-plane magnetization while still avoiding stray fields. This is demonstrated in Figure~\ref{fig:Figure4} for the three antiferromagnetic chains analyzed in Figure~\ref{fig:Figure4}, i.e. $[001]$ (spacing $a$ = 0.66 nm) , $[1\overline{1}0]$ ($a$ = 0.96 nm), and $[1\overline{1}1]$ ($a$ = 0.86 nm). Measurements have been acquired with the very same tip used in Figure~\ref{fig:Figure3} and at the same scanning bias ($V$ = 0.33 mV) but under the influence of an out-of-plane magnetic field of strength $B$ = 2 T. The alternating contrast between adjacent adatoms visible in the d$I$/d$U$ maps and relative line profiles for chains oriented along $[001]$ (Fig.~\ref{fig:Figure4}\textcolor{red}{a}) and $[1\overline{1}0]$ (Fig.~\ref{fig:Figure4}\textcolor{red}{b}) directions confirms their antiferromagnetic state. In sharp contrast, the atomic wire oriented along the $[1\overline{1}1]$ direction (spacing $a$ = 0.86 nm) is now characterized by a constant signal across the chain. This observation is indicative of a magnetic-field driven transition from an antiferromagnetic to a ferromagnetic state, which is expected to occur once the Zeeman energy overcomes the exchange coupling $J$ between adatoms. These results demonstrate that the magnetic interaction between adatoms coupled along the $[1\overline{1}1]$ direction with spacing $a$ = 0.86 nm is weak while it is stronger for the $[001]$ and $[1\overline{1}0]$ chains. We note that this conclusion is further supported by the direct comparison of the zero bias anomaly characterizing adatoms coupled along distinct crystallographic directions. As discussed in Supplementary Information, a clear change of spectral lineshape with respect to the single Cr case is visible for adatoms coupled along the $[001]$ and $[1\overline{1}0]$ directions. In sharp contrast, the d$I$/d$U$ curve for adatoms coupled along the $[1\overline{1}1]$ direction at spacing $a$ = 0.86 nm perfectly overlaps the single Cr adatom spectrum, signaling a very weak exchange term $J$ which, in line with our magnetic characterization, can be easily overcome by a moderate Zeeman term \cite{KBL2021,doi:10.1126/sciadv.abi7291}. 

\newpage
\textbf{YSR tips with tilted sensitivity}.

\begin{figure}[!t]
    \centering
    \includegraphics[width=.45\textwidth,page=1]{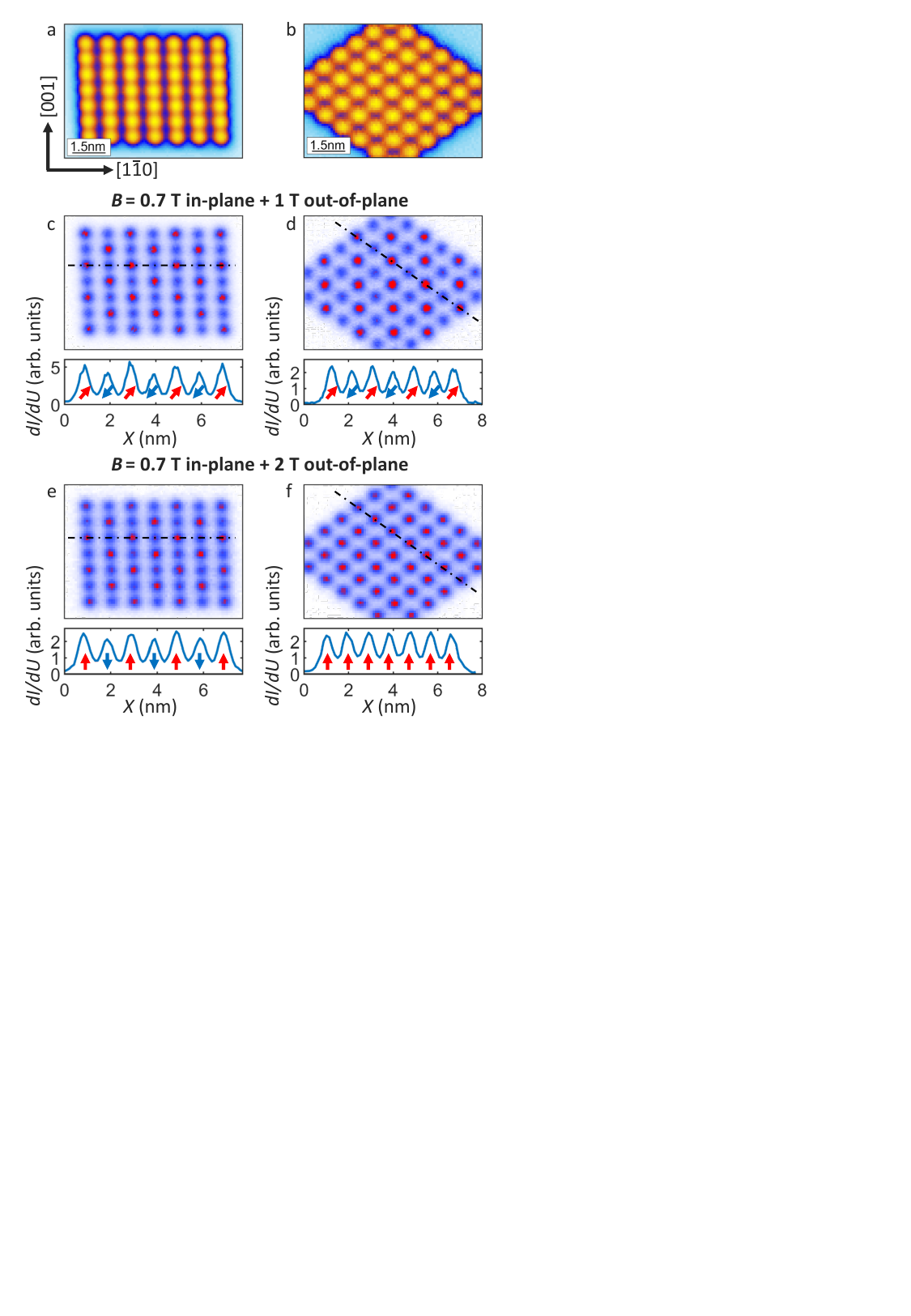}
    \caption{\textbf{Yu-Shiba-Rusinov tips for imaging 2D lattices}
    \textbf{a,b} Topographic images of two distinct 2D lattices assembled using Cr aadtoms coupled to the Nb(110) surface. \textbf{c,d} d$I$/d$U$ maps and corresponding line profiles acquired using a YSR tip spectroscopically similar to the one  discussed in Figure 2. Measurements have been performed under and out-of-plane magnetic field of strength $B$ = 1 T. Both lattice are characterized by alternating spin contrast indicative of antiferromagnetic coupling between nearest neighbour adatoms. \textbf{d,e} d$I$/d$U$ maps and corresponding line profiles acquired by increasing the magnetic field strength to $B$ = 2 T. While the rectangular lattice preserve its antiferromagnetic state, a transition to a ferromagnetic state is visible for the rhombic lattice. Stabilization parameters $-3~\mathrm{mV}$, $500~\mathrm{pA}$, scanning bias $0.33~\mathrm{mV}$, modulation $80~\mathrm{\upmu V}$.}
    \label{fig:Figure5}
\end{figure}

  The possibility of performing measurements in relatively high magnetic fields while preserving the spin sensitivity of the YSR state is further demonstrated in Figure~\ref{fig:Figure5}, which directly compares two distinct structures assembled atom-by-atom corresponding to a rectangular (Fig.~\ref{fig:Figure5}\textcolor{red}{a}) and a rhombic  (Fig.~\ref{fig:Figure5}\textcolor{red}{b}) lattice \cite{SKW2023}. The spin-resolved d$I$/d$U$ maps reported in Fig.~\ref{fig:Figure5}\textcolor{red}{c,d} and acquired under the influence of a tilted magnetic field ($B$ = 0.7 T in-plane, $B$ = 1 T out-of-plane) reveal the existence of an antiferromagnetic state in both structures, additionally highlighted by the alternating contrast visible in the respective line profiles. By further increasing the out-of-plane magnetic field to $B$ = 2 T while maintaining an in-plane field of 0.7 T, the rhombic lattice is driven into a ferromagnetic structure while the the rectangular structure retains its antiferromagnetic state. Both observations are in line with the results obtained on 1D wires discussed in Fig.~\ref{fig:Figure4}. We would like to note that all measurements have been performed using the very same YSR tip. This allows to exclude possible tip-related artifacts such as the loss of magnetic sensitivity which could explain the change of contrast observed in the rhombic structure, directly proving how YSR tips can be used to image magnetic-field driven transitions in the magnetic state at the atomic scale.

\textbf{Conclusions.}

Overall, our measurements reveal that the robustness of the superconducting cluster at the apex of functionalized YSR tips allows to perform spin-resolved spectroscopic imaging in relatively high magnetic fields. We demonstrate that this property confers YSR tips the possibility to sense in-plane, out-of-plane, and tilted magnetizations as well as to visualize magnetic field-driven transitions of the magnetic texture. We envision that this methodology will be particularly useful to provide a full-magnetic characterization of non-collinear spin structures. 

 	\bibliography{./References.bib}

\end{document}